# Hierarchy of exchange interactions in the triangular-lattice spin-liquid YbMgGaO$_4$

Xinshu Zhang,[1] Fahad Mahmood,[1, *] Marcus Daum,[2] Zhiling Dun,[3] Joseph A. M. Paddison,[2, 4]
Nicholas J. Laurita,[1] Tao Hong,[5] Haidong Zhou,[3] N. P. Armitage,[1] and Martin Mourigal[2]

[1]*Institute for Quantum Matter, Department of Physics and Astronomy,
Johns Hopkins University, Baltimore, Maryland 21218, USA*
[2]*School of Physics, Georgia Institute of Technology, Atlanta, Georgia 30332, USA*
[3]*Department of Physics and Astronomy, University of Tennessee, Knoxville, Tennessee 37996, USA*
[4]*Churchill College, University of Cambridge, Storey's Way, Cambridge CB3 0DS, UK*
[5]*Quantum Condensed Matter Division, Oak Ridge National Laboratory, Oak Ridge, TN 37831, USA*
(Dated: January 30, 2018)

The spin-1/2 triangular lattice antiferromagnet YbMgGaO$_4$ has attracted recent attention as a quantum spin-liquid candidate with the possible presence of off-diagonal anisotropic exchange interactions induced by spin-orbit coupling. Whether a quantum spin-liquid is stabilized or not depends on the interplay of various exchange interactions with chemical disorder that is inherent to the layered structure of the compound. We combine time-domain terahertz spectroscopy and inelastic neutron scattering measurements in the field polarized state of YbMgGaO$_4$ to obtain better insight on its exchange interactions. Terahertz spectroscopy in this fashion functions as high-field electron spin resonance and probes the spin-wave excitations at the Brillouin zone center, ideally complementing neutron scattering. A global spin-wave fit to all our spectroscopic data at fields over 4 T, informed by the analysis of the terahertz spectroscopy linewidths, yields stringent constraints on $g$-factors and exchange interactions. Our results paint YbMgGaO$_4$ as an easy-plane XXZ antiferromagnet with the combined and necessary presence of sub-leading next-nearest neighbor and weak anisotropic off-diagonal nearest-neighbor interactions. Moreover, the obtained $g$-factors are substantially different from previous reports. This works establishes the hierarchy of exchange interactions in YbMgGaO$_4$ from high-field data alone and thus strongly constrains possible mechanisms responsible for the observed spin-liquid phenomenology.

PACS numbers: 75.10.Kt,75.30.Et

## I. Introduction

Quantum spin-liquids (QSL) are exotic states of matter in which spins are highly correlated but remain dynamic down to zero temperature due to strong quantum fluctuations [1, 2]. Many distinct QSL states have been proposed theoretically [3, 4] and classified according to their non-local (topological) properties [5]. Their detection, however, remains a central challenge for condensed matter physics [6], and relies on the presence of quantum entanglement in their ground-state and fractional quasi-particles in their excitation spectra. While the former can be checked by numerics [7], the latter can be experimentally detected by thermodynamic techniques [8–10] or spectroscopic probes such as inelastic neutron scattering [11–17] and electron-spin resonance [18–21].

In spite of these recent breakthroughs, the most celebrated flavor of a QSL remains the resonating valence-bond (RVB) state first proposed by Anderson in 1973 [22] for the spin-1/2 triangular-lattice Heisenberg antiferromagnet (TLHAF) and later extended to the square-lattice in the context of cuprate superconductors [23]. However, precise numerical calculations for the spin-1/2 TLHAF, which take into account nearest neighbor interactions only [24–26], indicate that quantum fluctuations are not enough to suppress magnetic ordering and the actual ground-state is a non-collinear long-range ordered spin structure. Experiments on various spin-$S$ triangular-lattice antiferromagnets have overwhelmingly confirmed this picture [27–30], with a few noteworthy exceptions [31–33]. Several perturbations from the pure TLHAF have been proposed to enhance quantum fluctuations: next-nearest neighbor interactions (NNN) [34–38], ring-exchange terms [39, 40] and anisotropic exchange [41–44], although it remains theoretically unclear if the latter mechanism alone can stabilize a QSL [45–47]. Nevertheless, anisotropic exchange interactions [48, 49] are known to generate new physics in rare-earth pyrochlores such as Yb$_2$Ti$_2$O$_7$ [48, 50–52], making it worthwhile to investigate their effect in other lattice geometries.

The newly discovered rare-earth triangular-lattice antiferromagnet YbMgGaO$_4$ [53, 54] appears to fulfill precisely this promise. The magnetic Yb$^{3+}$ ions carry effective spin-1/2 moments in a symmetry environment allowing anisotropic exchange interactions [41, 47] in the absence of antisymmetric (Dzyaloshinsky-Moriya) terms and magnetic defects, both of which are present in other two-dimensional QSL candidates such as Herbertsmithite [55–57]. The immediate availability of single crystals [54] uncovered a QSL phenomenology in YbMgGaO$_4$ characterized by the absence of spin ordering or freezing down to $T = 100$ mK in muon spin relaxation measurements [58], much lower than the Curie-Weiss temperature $\theta_W \approx -4$ K, and a power-law behavior for the magnetic specific heat at low temperatures [54, 59]. Perhaps the strongest evidence for a QSL in YbMgGaO$_4$ came from inelastic neutron scattering measurements in zero field that unraveled a broad continuum of magnetic excitations across the entire Brillouin-zone [60–62]. This continuum has been interpreted as composed of fractional excitations from a U(1) QSL state with spinon Fermi surface [42, 44, 61] or from a RVB-like state [62].

---

* fahad@jhu.edu

The absence of a magnetic contribution to the thermal conductivity [59], however, appears difficult to reconcile with the spinon Fermi surface interpretation. Additionally, the disordered occupancy of the inter-triangular layers by Mg$^{2+}$ and Ga$^{3+}$ ions [53] appears to affect profoundly the Yb$^{3+}$ ions with broadened crystal electric-field (CEF) levels [60, 63], a distribution of g-tensors [63], and a broadened magnetic excitation spectrum at high fields [60]. The impact of disorder on the YbMgGaO$_4$ exchange Hamiltonian and, therefore, whether the ground state is a QSL or not, remains an outstanding issue [47]. In fact, the nature of the dominant exchange interactions in YbMgGaO$_4$ is also controversial. While the overall planar anisotropy is clear from susceptibility measurements [54, 61], both antiferromagnetic next-nearest-neighbor terms ($J_2$) [60] and nearest-neighbor anisotropic off-diagonal exchanges (so-called $J_1^{\pm\pm}$ and $J_1^{z\pm}$) [43, 64] have been proposed as extensions from the XXZ model. Comprehensively determining the exchange interactions in YbMgGaO$_4$ is of fundamental importance in deciphering the nature of its ground state.

Here, we combine time-domain THz spectroscopy (TDTS) with inelastic neutron scattering (INS) measurements in the field-polarized state of YbMgGaO$_4$ to extract values of the exchange interactions from a global fit to spin-wave spectra measured for different field directions and scattering wave-vectors including the Brillouin-zone center (Γ-point). Previous high-field INS measurements were limited to wave-vectors around the antiferromagnetic zone boundary [60] while previous X-band electron spin resonance (ESR) measurements [54] were intrinsically limited to small fields ($\lesssim 0.4$ T) below the field-polarized regime. In contrast, high resolution TDTS (e.g. [65]) functioning as high field ESR on magnetic insulators allows an accurate determination of magnetic excitations in fields comparable to the saturation field in YbMgGaO$_4$ [54, 60]. Since the wavelength of THz radiation is much greater than lattice constants, TDTS measures excitations in the first Brillouin zone with zero-momentum transfer (i.e. the $\mathbf{q}=0$ response), which is impossible with neutron scattering. Inclusion of the high-field TDTS data allows a substantial refinement of the Hamiltonian parameters compared to Ref. [60]. Moreover, an analysis of TDTS linewidths further constrains the anisotropic exchange interactions. In the context of prevailing models our results suggest that both NNN and off-diagonal anisotropic exchange interactions are present in YbMgGaO$_4$ — with pseudo-dipolar terms sub-leading compared with the XXZ part of the model — and that this nominal set of exchanges may lie closer to a spin-liquid regime than previously thought [47, 60]. When combined with the likely presence of exchange disorder, the extent to which remains to be determined, our work strongly constrains the nature of the YbMgGaO$_4$ ground state and points towards a more intricate underlying scenario than originally reported [61].

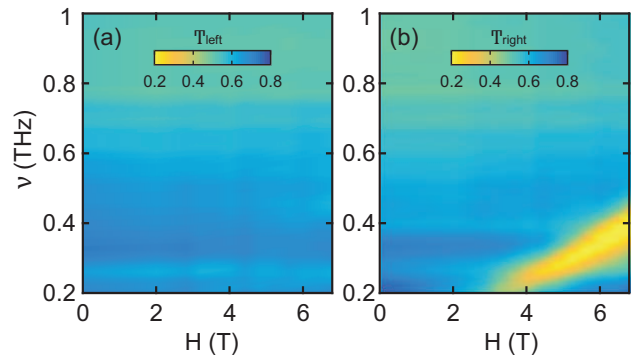

FIG. 1. Transmission amplitude for (a) left- and (b) right-circularly polarized THz light as a function of frequency $\nu$ and field $H_{dc}$ in the Faraday geometry, i.e. $H_{dc} \parallel c$ at $T=5$ K. The transmission curves were measured at 0 T, 1 T, 2 T and then at every 0.4 T from 2.8 T to 6.8 T. The bright yellow feature in (b) indicates the absorption due to the $\mathbf{q}=0$ excitation.

## II. Methods

### A. Experimental methods

The YbMgGaO$_4$ crystal (space group $R\overline{3}m$) used in this work was grown by the floating-zone technique (see Supplementary Information (SI) Section 1 for details) as reported in Ref. [60] and cut with a diamond blade to present a c-axis facet, where the c-axis is orthogonal to the triangular ab plane of Yb$^{3+}$ ions. TDTS measurements were performed in a custom built polarization modulation setup with a frequency range of 0.2 to 2 THz (0.83 to 8.3 meV) (see SI Sec. 2). The complex THz transmission of a $4 \times 4 \times 1$ mm$^3$ single crystal was measured down to 1.6 K (see SI Sec. 7) with external fields up to $\mathbf{H} = 6.8$ T in both the Faraday ($\mathbf{k} \parallel \mathbf{H}$) and Voigt ($\mathbf{k} \perp \mathbf{H}$) geometries, where $\mathbf{k}$ is the direction of light propagation. In both cases, the THz pulse ac magnetic field, $\mathbf{h}$, was applied along $\mathbf{a}^*$ with $\mathbf{h} \perp \mathbf{H}$ and thus the $\mathbf{H} \parallel \mathbf{c}$ and $\mathbf{H} \parallel \mathbf{a}$ orientations were probed in Faraday and Voigt geometries, respectively.

INS experiments were performed on the cold neutron chopper spectrometer [66] (CNCS) at the Spallation Neutron Source (SNS), and on the cold triple-axis spectrometer CTAX at the High Flux Isotope Reactor (HFIR), both at Oak Ridge National Laboratory. Experiments were performed on the same crystals as Ref. [60] with a magnetic field of $H = 7.8$ T along the crystal c-axis and $T \approx 0.06$ K on CNCS, and at $H = 10.8$ T along the crystal a-axis and $T \approx 0.32$ K on CTAX. The INS data taken at CNCS is the same as published in [60] but now analyzed together with the TDTS data. Given the broad spectra, even at high-fields, the energy of magnetic excitations for a given wave-vector was determined by fitting the maximum in scattering intensity (Ref. [60] and SI Sec. 3).

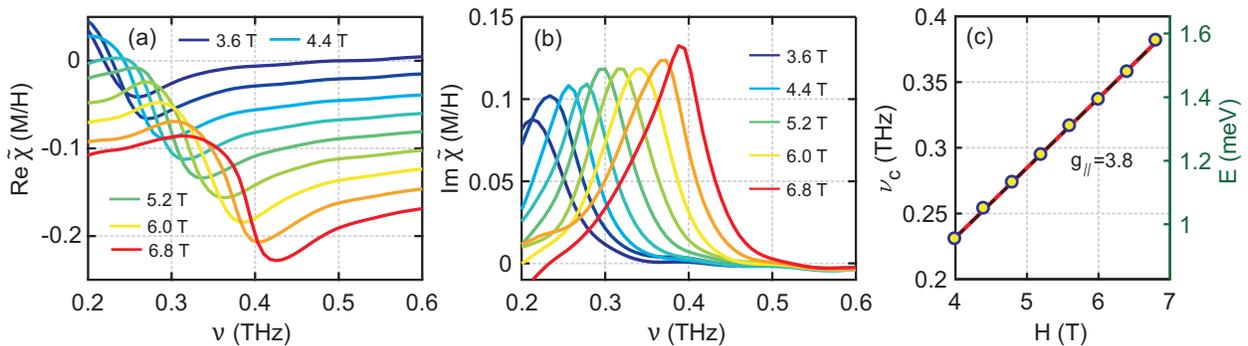

FIG. 2. Real **(a)** and imaginary **(b)** parts of the magnetic susceptibility $\tilde{\chi}(\nu)$ for right-circularly polarized light at different fields $H_{dc} \parallel c$ at 5 K. $\tilde{\chi}(\nu)$ is obtained by referencing the TDTS data to the spectra at 100 K. Spectra in (a) are offset vertically by 0.02 for clarity. **(c)** Resonant frequency ($\nu_c$) at 5 K vs magnetic field for magnetic excitation peaks in (b) in the Faraday geometry $H_{dc} \parallel c$. Yellow circles represent the data. Red line is a linear best fit to extract $g_\parallel$. The black dashed line represents the global fit to the data (see Fig 3). Error bars in **(c)** are smaller than the marker size.

### B. Analysis of the TDTS data

In the field-polarized state in the Faraday geometry, the linearly polarized THz pulse becomes elliptically polarized as it passes through the sample due to spins precessing around the applied field direction. The complex transmission is represented by a $2 \times 2$ Jones matrix. Due to the three-fold rotational symmetry of the lattice, this reduces to an anti-symmetric matrix for the transmission of a linearly polarized pulse [67]. By diagonalizing this antisymmetric matrix, we can convert the transmission matrix from the linear basis into a circular basis (see SI Sec. 2) which naturally corresponds to eigenstates of the transmission in the Faraday geometry [50]. In this manner we can separate the transmission of a left (LCP) and right (RCP) circularly polarized THz pulse.

Fig. 1a and Fig. 1b show such transmission magnitudes for LCP and RCP THz light, respectively, and as a function of frequency and magnetic field at a temperature of $T = 5$ K. The bright yellow feature in Fig. 1b shows that only one particular helicity of the light (RCP) is strongly absorbed. This indicates the presence of well defined spin-wave excitations with energy linearly dependent on the applied field. In a field polarized regime in the Faraday geometry, the direction of spin precession is determined by the orientation of the applied $H_{dc}$ field and it is therefore natural to expect an absorption for only one helicity and not the other. Indeed, our experiments confirm that when the polarity of the $H_{dc}$ field is flipped, only LCP light is absorbed (see SI Sec. 4). This confirms that the absorption line observed in Fig. 1b originates from spin-wave excitations in the field-polarized regime of YbMgGaO$_4$. The observed absorption is the only feature in the data that is affected by the $H_{dc}$ field and temperature, suggesting the featureless background has a non-magnetic origin and is likely the low-energy tail of a crystal field level absorption (see SI Sec. 5).

The featureless nature of the background allows us to extract the frequency dependent magnetic susceptibility $\tilde{\chi}(\nu)$ as follows. The complex transmission of a particular eigenpolarization (LCP or RCP light) is given by the relation $\tilde{T}(\nu) = [4\tilde{n}/(\tilde{n}+1)^2]\exp[i2\pi \nu d(\tilde{n}-1)/c]$, where $\tilde{n} = \sqrt{\tilde{\epsilon}\tilde{\mu}}$ is the complex index of refraction, $\tilde{\epsilon}$ is the dielectric constant, $\tilde{\mu} = 1 + \tilde{\chi}$ and $d$ is the sample thickness. We determine $\tilde{n}$ using the Newton-Raphson method and then isolate $\tilde{\chi}$ by measuring the sample at a reference temperature at which the spectrum does not show any signatures of the spin-wave absorption [51]. At this reference temperature (100 K in this case, see SI Sec. 6) $\tilde{\chi}$ can be taken to be zero and so $\tilde{n}_{100K} = \sqrt{\tilde{\epsilon}}$. Thus, the low temperature magnetic susceptibility is given by $\tilde{\chi} = (\tilde{n}_{5K}/\tilde{n}_{100K})^2 - 1$ [68].

Fig. 2a-b show the real and imaginary parts of $\tilde{\chi}(\nu)$, respectively, at different fields and $T = 5$ K in the Faraday geometry. Peaks in Im$\tilde{\chi}(\nu)$ correspond to the spin wave excitations in the $\mathbf{q} = 0$ limit. By fitting the data at each field with a Lorentzian (see SI Sec. 7) we extract the resonant energy ($E$) of the spin-wave absorption. The resulting $E$ vs $H_{dc}$ plot is shown in Fig. 2c. The peak widths and resonant frequencies show little temperature dependence between 1.6 K and 40 K (see SI Sec. 7). Similar analysis (see SI Sec. 8) is done for TDTS in the Voigt geometry to extract the spin-wave energies for field $H_{dc}$ along the $a$-axis (reported in Fig. 3). This TDTS data is then combined with INS data to extract the exchange interactions in a global fit, as discussed below.

### C. High-field spin-wave theory analysis

An effective spin-1/2 Hamiltonian relevant for YbMgGaO$_4$ has been given by [41, 42, 54, 60]:

$$\mathcal{H} = \sum_{\langle i,j \rangle} [J_1^{zz} S_i^z S_j^z + J_1^{\pm}(S_i^+ S_j^- + S_i^- S_j^+)$$
$$+ J_1^{\pm\pm}(\gamma_{ij} S_i^+ S_j^+ + \gamma_{ij}^* S_i^- S_j^-)$$
$$- \frac{iJ_1^{z\pm}}{2}(\gamma_{ij}^* S_i^+ S_j^z - \gamma_{ij} S_i^- S_j^z + \langle i \leftrightarrow j \rangle)]$$
$$+ \sum_{\langle\langle i,j \rangle\rangle} [J_2^{zz} S_i^z S_j^z + J_2^{\pm}(S_i^+ S_j^- + S_i^- S_j^+)]$$
$$- \mu_0 \mu_B \sum_i [g_\perp(H^x S_i^x + H^y S_i^y) + g_\parallel H^z S_i^z] \quad (1)$$

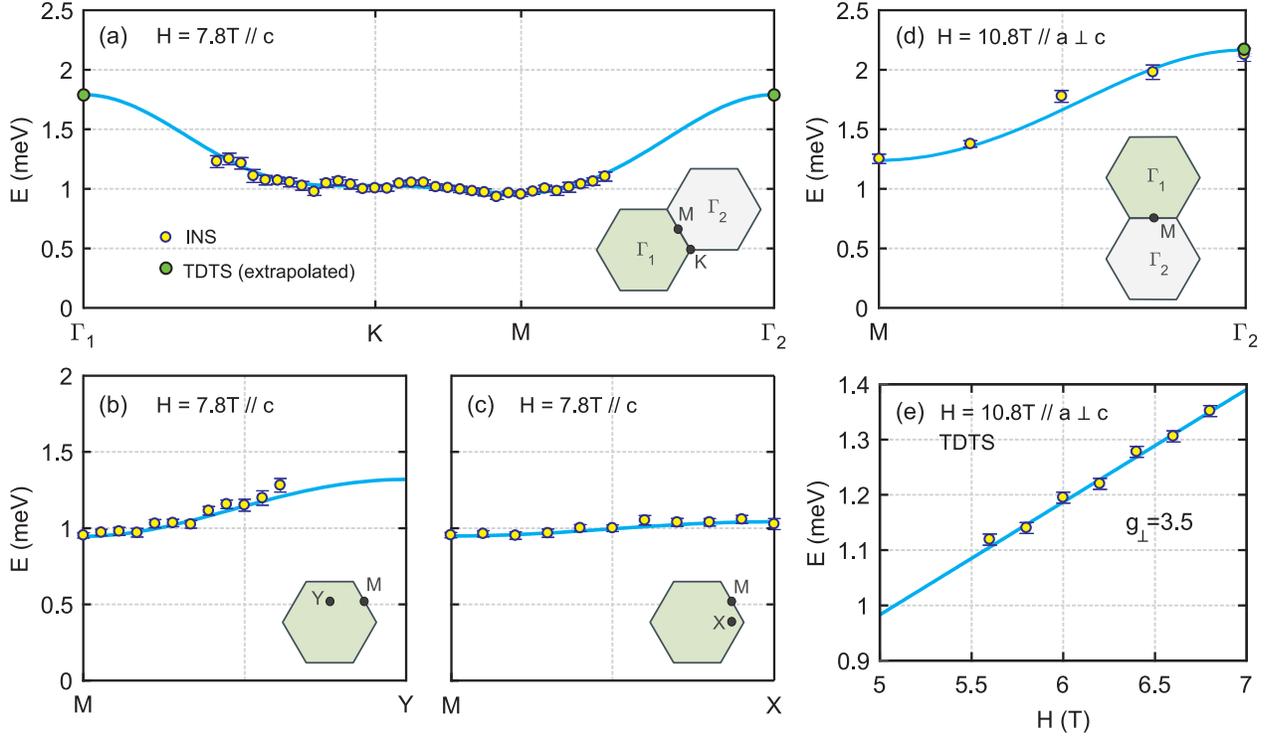

FIG. 3. **(a)-(c)** Energy dependence of spin-wave excitations in the field-saturated state from INS data along high symmetry directions of the triangular Brillouin zone. The sample was cooled by a dilution refrigerator of base temperature $T = 0.06$ K in a magnetic field $H_{dc} \parallel c = 7.8$ T. Solid blue lines show the results of the global fit using model C (see Table I). Green circles at the $\Gamma$-points are the TDTS data from Fig. 2c extrapolated to 7.8 T and included for display purposes. Errorbars for this extrapolated data are smaller than the marker size. **(d)** INS data measured at $T = 0.32$ K in a field $H_{dc} \parallel a = 10.8$ T. Yellow circles show the position of the maximum intensity at each wave-vector as determined from fits to the energy dependent data. Green circle at the $\Gamma_2$ point shows the TDTS data extrapolated to 10.8 T. **(e)** Energy vs. field for spin-wave excitations at $T = 5$ K in the TDTS Voigt geometry, i.e. $H_{dc} \parallel a$. Yellow circles represent the resonant energy of the spin-wave peaks in the TDTS spectra (see SI). Solid red lines in (d) and (e) show the results of the global fit using model C (Tab. I). Note that the INS data in (a)-(c) is the same as published in [60].

where $S^{\pm} = S^x \pm iS^y$, $g_\parallel$ and $g_\perp$ are components of the g-tensor parallel and perpendicular to the c-axis, complex numbers $\gamma_{ij}$ are defined in [54] and brackets $<>$ and $\ll\gg$ represent nearest and next-nearest neighbor pairs respectively. The exchange interactions $J^{zz}$ and $J^{\pm}$ originate from the standard XXZ model, with subscripts 1 and 2 indicating the nearest and next-nearest neighbor interactions respectively. We also include symmetry allowed bond-dependent interactions $J^{\pm\pm}$ and $J^{z\pm}$ (also known as pseudo-dipolar interactions), which have a spin-orbit origin. The pseudo-dipolar interactions between next-nearest neighbors are neglected since they are supposed to be small [69, 70] . For large applied fields and assuming a field-polarized state, the above Hamiltonian can be solved to yield the spin wave dispersions for external fields along the c- and a-axes (see SI).

By setting $\mathbf{q} = 0$, the spin wave excitations for the Faraday and Voigt geometry can be simplified to:

$$E(H^z) = \mu_0\mu_B g_\parallel H^z - 3(J_1^{zz} + J_2^{zz}) + 6(J_1^{\pm} + J_2^{\pm}) \quad (2)$$

$$E(H^x) = \{[\mu_0\mu_B g_\perp H^x + \frac{3}{2}(J_1^{zz} + J_2^{zz}) - 3(J_1^{\pm} + J_2^{\pm})]^2$$
$$- |\frac{3}{2}(J_1^{zz} + J_2^{zz}) - 3(J_1^{\pm} + J_2^{\pm})|^2\}^{\frac{1}{2}}. \quad (3)$$

According to Eq. 2, the field dependence in the Faraday geometry is particularly simple and depends only on $g_\parallel$. From a simple linear fit to the data in Fig. 2c we obtain $g_\parallel = 3.81(4)$. Note that the effective g-tensor here is a property of a single $Yb^{3+}$ ion and as such is independent of the exchange interactions [48]. This allows us to treat $g_\parallel$ as a fixed parameter in the subsequent global fit. In contrast, $E(H^x)$ is not linearly related to $g_\perp$ (Eq. 3) and so $g_\perp$ must be extracted simultaneously with the exchange constants of Eq. 1. One expects the present measurements of the g-factor to be considerably more accurate than previous X-band ESR results [54] due to greater than a ten-fold increase in the magnitude of the magnetic fields. The expressions for the dispersions with $\mathbf{q} \neq 0$, as relevant for neutron-scattering measurements, are given in Refs. [43, 60] and SI Sec. 3.

### III. Results

#### A. Fits to spin-waves energies

We now turn to our central result, which is to refine the parameters of $YbMgGaO_4$ by combining all the data at hand. In Fig. 3, we show the spin wave dispersion ob-tained through INS [Fig. 3(**a-d**)] as well as the TDTS data in Voigt geometry [Fig. 3(**e**)]. The entire dataset is fit simultaneously to the spin-wave dispersions in the field-saturated state obtained from the Hamiltonian in Eq. 1. There are six target parameters for this global fit, which are $J_1^{zz}$, $J_1^{\pm}$, $J_1^{\pm\pm}$, $|J_1^{z\pm}|$, the ratio $J_2/J_1$ and $g_\perp$ with a fixed $g_\parallel = 3.81$. Note that only the magnitude of $J_1^{z\pm}$ can be determined from the spin-wave dispersions (see SI Sec. 9). The signs of the other $J$ terms obtained from the fit is verified by starting from both negative and positive initial values. As spin-space anisotropy is primarily a property of the effective spin-1/2 doublet of $Yb^{3+}$, we adopt the same overall XXZ exchange anisotropy for both nearest neighbor and next-nearest neighbor interactions, i.e. $J_2^{zz}/J_2^{\pm} = J_1^{zz}/J_1^{\pm}$. This reasonable assumption helps in reducing the size of the parameter space for the global fit.

We obtain an excellent fit to the data using the above model [see lines in Fig. 3 and spread in Fig. 4] and report our global fitting parameters in Tab. I (Model C). The $\chi^2$ goodness of fit and correlation plots of these parameters are shown in the SI Sec. 10. We note that the above fit is performed with the constraint $|J_1^{zz}| > |J_1^{z\pm}|$. Without this, we obtain $|J_1^{z\pm}| = 0.45$ which we regard as unphysical as it is nearly three times larger than $J_1^{zz}$ and also considerably outside the bounds set by the linewidth analysis performed below.

In previous works, two different sets of parameters have been proposed for the exchange interactions of YbMgGaO$_4$. The first set [60], model A in Tab. I, includes both nearest and next-nearest neighbor interactions but sets $J_1^{\pm} = 0$ based on previous X-band ESR results [54]. The second set [43], model B in Tab. I, includes only nearest-neighbor interactions, with the $J_2 = 0$ constraint argued as a consequence of the localized nature of $Yb^{3+}$ 4f electrons [43, 64]. Both models used the $g$-factors determined from low frequency X-band ESR. Model C is the new best fit from our spin-wave analysis. To compare these models, we plot in Fig. 4 the experimental vs. calculated spin-wave energies for each of model A, B and C. Clearly, the points obtained for our model C lie significantly closer to the $E_{exp} = E_{calc}$ line than models A and B, highlighting the advantage of our enriched datasets over previous works.

Our global analysis yields finite NNN interactions with $J_2/J_1 = 0.18(7)$ and a slightly sub-dominant pseudo-dipolar $J_1^{\pm\pm} \approx 0.07$ meV interaction. The large fitting error bars on some parameters of the model reveal the high degree of correlation between $J_2$ and $J_1^{\pm\pm}$ and the very weak sensitivity of our fit to $|J_1^{z\pm}|$. This is apparent in the $\chi^2$ goodness of fit 2D plots (see SI Sec. 10) in the parameter spaces of $J_1^{\pm\pm}$ versus $|J_1^{z\pm}|$, and $J_1^{\pm\pm}$ versus $J_2/J_1$, respectively. The poor sensitivity to $|J_1^{z\pm}|$ is also clear from the analytical expression for the field-polarized spin-wave dispersion with field along the $a$-axis (see SI Sec. 9).

Given the correlation between $J_1^{\pm\pm}$ and $J_2$, it is natural to analyze how our results change by enforcing the constraint $J_2 = 0$ while leaving all other parameters free. This leads to model B*, which resembles model C except for a larger $J_1^{\pm\pm} = 0.13(2)$ meV which is now comparable to the dominant $J_1^{zz}$ exchange (see Tab. I). This model gives only a slightly worse fit to the spin-wave data than model C, as can be seen from a plot of the experimental against calculated spin-wave energies in Fig. 4. While a unique set of exchange parameters cannot be obtained from the above analysis, a definitive hierarchy of interactions nevertheless emerges. It yields easy-plane XXZ terms for YbMgGaO$_4$ with $J_1^{zz} = 0.15(1)$ meV, $\Delta = J_1^{zz}/2J_1^{\pm} = 0.9(1)$, a subleading pseudo-dipolar term $J_1^{\pm\pm}/J_1^{zz} \leq 1$ and a small NNN exchange $J_2^{zz} = 0.03(1)$ meV.

| Model | A | B | B* | C |
|---|---|---|---|---|
| $J_1^{zz}$ (meV) | 0.126 | 0.164 | 0.151(5) | 0.149(5) |
| $J_1^{\pm}$ (meV) | 0.109 | 0.108 | 0.088(3) | 0.085(3) |
| $J_1^{\pm\pm}$ (meV) | 0.013 | 0.056 | 0.13(2) | 0.07(6) |
| $|J_1^{z\pm}|$ (meV) | 0 | 0.098 | 0.1(1) | 0.1(1) |
| $J_2/J_1$ | 0.22 | 0 | 0 | 0.18(7) |
| $g_\parallel$ | 3.72 | 3.72 | 3.81(4) | 3.81(4) |
| $g_\perp$ | 3.06 | 3.06 | 3.53(5) | 3.53(5) |

TABLE I. Exchange parameters for different models derived from fitting the spin-wave dispersions. Models A and B are from [60] and [43], respectively. Model C is from our global fit to the TDTS and INS data. Model B* is from a global fit to the data by ignoring NNN interactions, i.e. $J_2 = 0$. Uncertainties in the values represent the 99.7% confidence interval (3 s.d.) in extracting the fitting parameters.

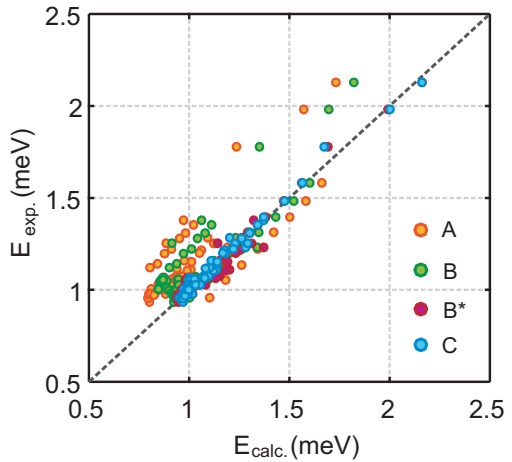

FIG. 4. Experimental vs. calculated spin-wave energies, where the dashed line denotes $E_{exp} = E_{calc}$. Fitted exchange parameters for models A, B, and C are listed in I. Models A (Ref. [60]), B (Ref. [43]), and C (global fit to the TDTS and THz data) are represented by orange, green and blue symbols, respectively. Model B* (global fit to data under the $J_2 = 0$ constraint) is represented by pink symbols.

### B. Further constraints: TDTS lineshapes

To further constrain these parameters, we require additional information. For that, we analyzed the linewidths of the spin-wave peaks in the TDTS data of Fig. 2. As noted above, TDTS functions here as high-field ESR [56]. ESR linewidths are strongly sensitive to the magnitude of the off-diagonal anisotropic interactions $J_1^{\pm\pm}$ and $J_1^{z\pm}$ of our model. The analysis proceeds using the formalism of Kubo-Tomita [54, 71, 72] which relates the width of the ESR absorption line to commutators that depend only on the anisotropic exchange interactions. This perturbative result is valid for small anisotropic exchanges when temperature and magnetic field are larger than the dominant exchange, conditions that are realized in our high-field TDTS experiments. The analysis [54] shows the Lorentzian width $\Delta_{cal} = \sqrt{2\pi M_2^3/M_4}$, where $M_2$ and $M_4$ are the second and fourth moment of the anisotropic part of the Hamiltonian (see SI Sec. 11).

Here we use this formalism to show that the observed spin-wave linewidths cannot be reconciled with the large anisotropic exchange terms suggested by model B* and therefore $J_2$ must be non-zero. In Fig. 5, we plot the deviation between the calculated and the experimental spin-wave resonance linewidths in both the Faraday $\Delta_f$ and Voigt $\Delta_v$ geometry as a function of $|J_1^{z\pm}|$ and $|J_1^{\pm\pm}|$. We analyze the case of $J_2 = 0$ that is relevant for model B*. We define a function $R_p = \frac{1}{2}[|\frac{\Delta_f - \Delta_{cal}}{\Delta_f}| + |\frac{\Delta_v - \Delta_{cal}}{\Delta_v}|]$, that is minimal along the yellow contour in Fig. 5, for which a rough analytical description is $\sqrt{|J_1^{z\pm}|^2 + |J_1^{\pm\pm}|^2} \approx 0.04(1)\,\text{meV}$. This calculation assumes that all the broadening of the THz data comes from exchange anisotropy with no distribution of $g$-factors beyond the values determined above. In fact, a relatively large spread in the latter values, $\Delta g_\parallel/\bar{g}_\parallel \approx 0.3$ and $\Delta g_\perp/\bar{g}_\perp \approx 0.1$, has been demonstrated by a careful analysis of CEF excitations linewidths [63] and must also play a role in the lineshapes observed here. Our analysis of the high-field TDTS lineshapes therefore provides an upper bound on any off-diagonal anisotropic exchanges $|J_1^{z\pm}|$ and $|J_1^{\pm\pm}|$. The parameters determined in model B* are incompatible with this bound and so model B* is ruled out as a realistic description of the exchange parameters of YbMgGaO$_4$.

Taken together, our results strongly favor an easy-plane XXZ scenario for YbMgGaO$_4$ with the combination of finite $J_2$ and relatively small pseudo-dipolar exchanges. Thus, the refinement of model C with a range of best possible parameter values we found for YbMgGaO$_4$ can be summarized as a model C*:

$$J_1^{zz} = 0.149(5)\text{ meV}, \quad J_1^{\pm} = 0.085(3)\text{ meV},$$

$$J_2/J_1 = 0.18(7), \quad \sqrt{|J_1^{\pm\pm}|^2 + |J_1^{z\pm}|^2} \lesssim 0.05\text{ meV}.$$

The above analysis assumes that finite $J_2$ will also give an additional contribution to the linewidth and hence the bounds on $|J_1^{\pm\pm}|$ and $|J_1^{z\pm}|$ for $J_2 = 0$ represent the maximum values that these parameters can take. We note that the bound determined for $|J_1^{z\pm}|$ justifies the earlier constraint of $|J_1^{zz}| > |J_1^{z\pm}|$ in determining the parameters for model C in Table I.

### C. Discussion

Experiments using other techniques, such as unpolarized [60] and polarized [64] neutron diffraction in zero field and low-temperatures may help to constrain the values further. Indeed, classical Monte-Carlo simulations for the instantaneous spin structure-factor $S(\mathbf{Q})$ reveal that either large $J_1^{\pm\pm}$ or relatively small $J_2 \sim 0.2 J_1$ yield correlations peaked at the M-point of the triangular-lattice Brillouin zone [60]. Maximal correlations at the M-point have been reported in all neutron scattering investigations to date [60–62, 64]. In this regard, we note that Monte-Carlo simulations with a large $J_1^{\pm\pm}$ yield a strong modulation in $S(\mathbf{Q})$ from the first to the second Brillouin zone [60]. To the best of our knowledge, such a modulation has not yet been observed experimentally, which appears consistent with the relatively small $J_1^{\pm\pm}$ term indicated by our spin-wave resonance linewidth analysis. In general, the sensitivity of $S(\mathbf{Q})$ to spin anisotropy arises because neutrons scatter only from spin components perpendicular to $\mathbf{Q}$. However, this projection has not always been fully included in theoretical calculations of the scattering pattern, which may help to account for the somewhat larger values of $J_1^{\pm\pm}$ proposed in Refs. [43, 64].

Future magnetization studies at very low temperatures will be another important test for the presence of further-neighbor or anisotropic exchange interactions in YbMgGaO$_4$. The presence of multiple phase transitions and magnetization plateaus in a moderate magnetic field [73] is expected in the case of a XXZ triangular-lattice antiferromagnet [74, 75]. This may also help to constrain the possible values of $J_2$ [76] and anisotropic exchanges, although exchange disorder may be very efficient at suppressing these features [77]. Due to their non-spin-conserving nature, dominant pseudo-dipolar interactions should manifest as an asymptotic approach to saturated magnetization, as recently observed in the candidate Kitaev spin-liquid $\alpha$-RuCl$_3$ [78]. Finally, we note that the role and extent of CEF and exchange disorder remains an outstanding issue in YbMgGaO$_4$ with possible impact ranging from the mimicry of a spin-liquid [47] to disorder-induced entanglement [79, 80].

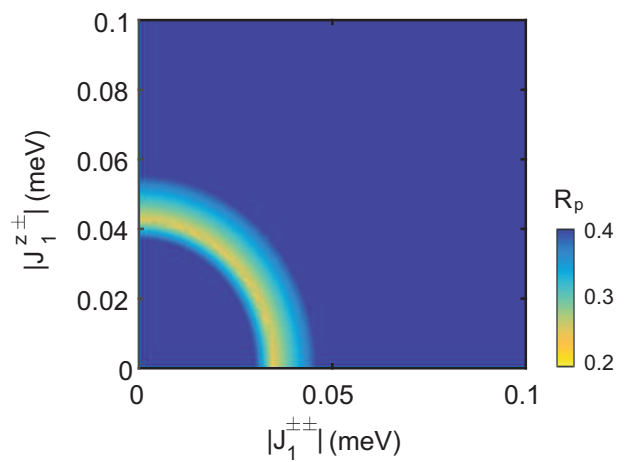

FIG. 5. Deviation $R_p$ of the experimental TDTS spin-wave resonance linewidths from the theoretically calculated ones as a function of the pseudo-dipolar interactions $|J_1^{z\pm}|$ and $|J_1^{\pm\pm}|$. This sets an upper bound on the pseudo-dipolar interactions in a model with $J_2 = 0$.

## IV. Conclusion

In conclusion, combining time-domain terahertz spectroscopy and neutron scattering in the high-field regime of YbMgGaO$_4$ yields the strongest constraints to date on possible mechanisms for the observed spin-liquid phenomenology. We note that a QSL regime is predicted for the spin-1/2 $J_1$-$J_2$ Heisenberg triangular-lattice antiferromagnet for $0.06 \lesssim J_2/J_1 \lesssim 0.19$ [36–38]. While a previous analysis obtained $J_2/J_1 = 0.22$ [60], the present work yields a slightly smaller $J_2/J_1 = 0.18(7)$. Additionally, the ratio $J_1^{zz}/J_1^{\pm} = 1.75(5)$ suggests that YbMgGaO$_4$ is more spin-isotropic than previously thought ($J_1^{zz}/J_1^{\pm} = 1.16$ [60]) with $J_1^{zz}/J_1^{\pm} = 2$ and 0 the Heisenberg and XY limits, respectively. We note, however, that subleading pseudo-dipolar interactions are also necessary to best explain our data. When combined with the likely presence of exchange disorder due to Mg$^{2+}$ and Ga$^{3+}$ disorder, this makes the spin-liquid mimicry [47] or the $J_1$-$J_2$ quantum spin-liquid [36–38] mechanisms serious contenders to the various scenarios proposed to explain the physics of YbMgGaO$_4$ thus far [44, 61, 62].

### Acknowledgments

The authors would like to thank C. Batista, C. Broholm, H. Changlani, G. Chen, S. Chernyshev, S. Haravifard, J. Rau, and O. Starykh for helpful discussions. Research at JHU was funded by the US DOE, Office of Basic Energy Sciences, Division of Materials Sciences and Engineering through Grant No. DE-FG02-08ER46544. Research at GT was supported by the College of Sciences and a Ralph E. Powe Junior Faculty Enhancement Award. The work at the University of Tennessee (Z.L.D. and H.D.Z.) was supported by the National Science Foundation through award DMR-1350002. The research at Oak Ridge National Laboratory's Spallation Neutron Source and High Flux Isotope Reactor was sponsored by the U.S. Department of Energy, Office of Basic Energy Sciences, Scientific User Facilities Division.

# Hierarchy of exchange interactions in the triangular-lattice spin-liquid YbMgGaO$_4$

Xinshu Zhang,[1] Fahad Mahmood,[1, *] Marcus Daum,[2] Zhiling Dun,[3] Joseph A. M. Paddison,[2, 4]
Nicholas J. Laurita,[1] Tao Hong,[5] Haidong Zhou,[3] N. P. Armitage,[1] and Martin Mourigal[2]

[1]*Institute for Quantum Matter, Department of Physics and Astronomy,
Johns Hopkins University, Baltimore, Maryland 21218, USA*
[2]*School of Physics, Georgia Institute of Technology, Atlanta, Georgia 30332, USA*
[3]*Department of Physics and Astronomy, University of Tennessee, Knoxville, Tennessee 37996, USA*
[4]*Churchill College, University of Cambridge, Storey's Way, Cambridge CB3 0DS, UK*
[5]*Quantum Condensed Matter Division, Oak Ridge National Laboratory, Oak Ridge, TN 37831, USA*
(Dated: January 29, 2018)

## 1. Sample Preparation

Polycrystalline samples of YbMgGaO$_4$ were synthesized by a solid state method. Stoichiometric ratios of Yb$_2$O$_3$, MgO, and Ga$_2$O$_3$ fine powder were carefully ground and reacted at a temperature of 1450 °C for 3 days with several intermediate grindings. Single-crystal samples of YbMgGaO$_4$ were grown using the optical floating-zone method under a 5 atm oxygen atmosphere. The best single crystals were obtained with a pulling speed of 1.5 mm/h, and showed [001] surfaces after several hours of growth.

## 2. Time-domain THz (TDTS) setup and analysis

Time-domain THz spectroscopy was performed in a home built setup at Johns Hopkins University. THz pulses, with a bandwidth between 0.2 to 2 THz, were generated by a photoconductive antenna (Auston switch) – emitter – upon illumination by an infrared laser and then detected by another Auston switch (receiver). The sample was mounted on a 4 mm aperture. The electric field profiles of the THz pulses transmitted through the sample and an identical bare aperture were recorded as a function of time and then converted to the frequency domain by Fast Fourier Transforms (FFTs). By dividing the FFTs of the sample and aperture scans, we can obtain the complex transmission of the sample.

In linear basis, the electric field vectors are represented as $E_i$ and $E_t$ for the incident and transmitted THz pulses respectively. Similarly, we define $E_{i,cir}$ and $E_{t,cir}$ in the circular basis. $\Lambda$ represents the transformation from circular to linear basis and we have $E_i = \Lambda E_{i,cir}$ and $E_t = \Lambda E_{t,cir}$ where $\Lambda = \frac{1}{\sqrt{2}}\begin{bmatrix} 1 & 1 \\ -i & i \end{bmatrix}$. The complex transmission of the sample in the linear basis is represented by a 2 × 2 Jones matrix: $T = \begin{bmatrix} T_{xx} & T_{xy} \\ T_{yx} & T_{yy} \end{bmatrix}$. The transmitted electric field through the sample is then given by: $E_t = TE_i$. For the circular basis, $E_{t,cir} = T_{cir}E_{i,cir}$, where $T_{cir} = \Lambda^{-1}T\Lambda$ is the transmission matrix in circular basis. The transmission matrix of a sample with three-fold rotational symmetry in Faraday geometry is fully antisymmetric [1], i.e. $T_{xx} = T_{yy}$ and $T_{xy} = -T_{yx}$. Therefore, the transmission matrix in the circular basis is diagonalized as $T_{cir} = \begin{bmatrix} T_{xx}+iT_{xy} & 0 \\ 0 & T_{xx}-iT_{xy} \end{bmatrix} = \begin{bmatrix} T_r & 0 \\ 0 & T_l \end{bmatrix}$, where $T_r$ and $T_l$ denote transmission matrix of right and left polarized light, respectively. With the polarization modulation technique as discussed in [2], $T_{xx}$ and $T_{xy}$ can be obtained simultaneously with a fast rotator. This allows us to determine two the complex transmission coefficients $T_{xx}$ and $T_{xy}$ simultaneously and convert these to $T_r$ and $T_l$.

## 3. Fitting INS data

The inelastic neutron scattering lineshapes obtained on CTAX are presented in Fig. S1 with the scattering wave-vector spanning the M$\Gamma_2$ direction with a magnetic field of 10.8 T applied along the $a$-axis of the crystal. The position of the most intense peak is indicated by vertical mark.

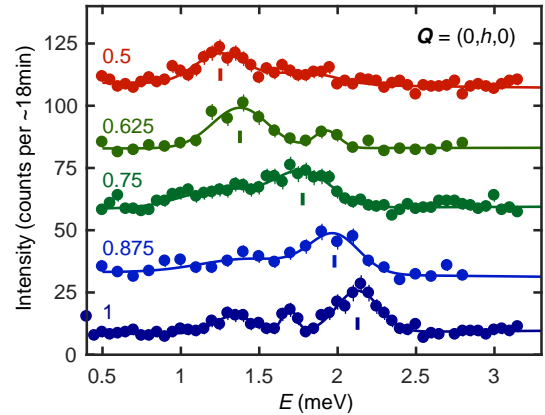

FIG. S1. Representative fit to the $E$-dependence of the INS data, used to determine the position of the intensity maximum at a given wave-vector **q**. This data was obtained for $H = 10.8\,\text{T} \parallel a$ at $T = 320\,\text{mK}$.

---

* fahad@jhu.edu

### 4. Faraday transmission for negative fields

To confirm the nature of the observed $YbMgGaO_4$ excitations in the Faraday geometry, we performed TDTS for both positive and negative magnetic field polarities. Here positive field refers to the field direction described in Fig. 1 of the main text. There the absorption appeared for only right circularly polarized (RCP) light and not for left circularly polarized (LCP) light. By reversing the field polarity, LCP light is now absorbed and not RCP light, as shown in Fig. S2. Compare this plot to Fig. 1 in the main text.

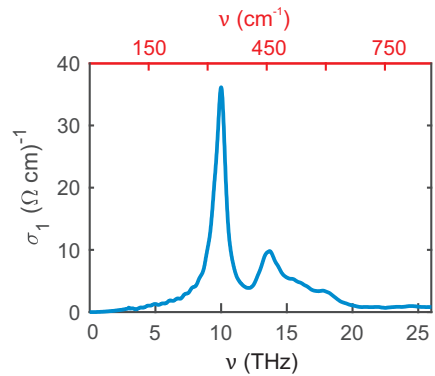

FIG. S3. Real part of the conductivity with frequency obtained from FTIR at 5 K.

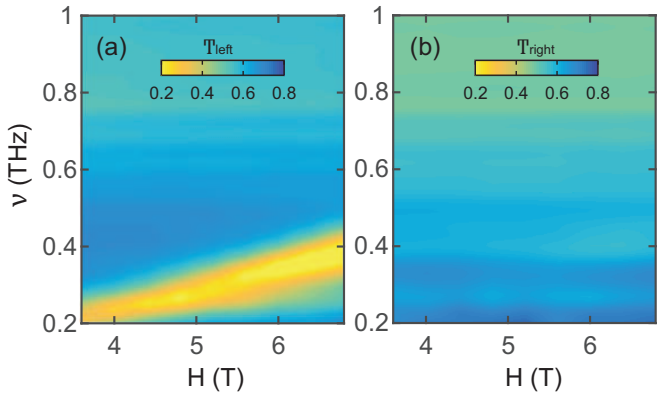

FIG. S2. Transmission amplitude for **(a)** left and **(b)** right circularly polarized light as a function of frequency and magnetic field strength in the Faraday geometry at $T = 5\,\mathrm{K}$.

### 5. FTIR spectra

We also performed Fourier transform infrared spectroscopy (FTIR) on the $YbMgGaO_4$ single crystal at $T = 5\,\mathrm{K}$. The reflectivity of the sample was measured using a commercial FTIR spectrometer in the far and mid infrared spectral ranges, i.e. $50\,\mathrm{cm}^{-1}$ to $7000\,\mathrm{cm}^{-1}$ (1.5 THz to 250 THz). The complex optical conductivity is calculated from the reflectivity spectrum using Kramers-Kronig constrained variation dielectric function fitting [3]. The real part of the optical conductivity $\sigma_1$ is shown in Fig. S3. The peaks in $\sigma_1$ match with the crystal field excitations observed by neutron scattering at 9.2 THz (38 meV) and 17.7 THz (61 meV) and 23.5 THz (97 meV). The low energy tail of the lowest energy crystal field excitation gives rise to the background observed in the TDTS measurements.

### 6. TDTS - Temperature dependence

In the main text we isolate the magnetic susceptibility $\tilde{\chi}(\nu)$ by measuring the sample at a reference temperature at which the spectrum does not show any signatures of the spin-wave absorption. We show the measured transmission spectrum in TDTS for the $YbMgGaO_4$ sample at different temperatures between $T = 5\,\mathrm{K}$ and $T = 100\,\mathrm{K}$ at $H = 6.8\,\mathrm{T}$ in the Faraday geometry (Fig. S4). The observed spin-wave absorption is the only feature in the data that is affected by the static field and temperature, suggesting that the featureless background has a non-magnetic origin. We do not observe any spin-wave absorption at $T = 100\,\mathrm{K}$ which allows us to use this as the reference temperature.

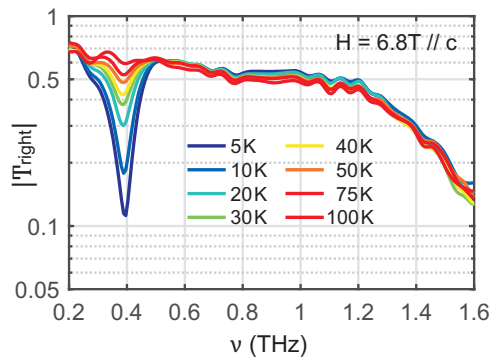

FIG. S4. Transmission amplitude for RCP light in the Faraday geometry taken from 5 K to 100 K at +6.8 T.

We also measured $YbMgGaO_4$ at $T = 1.6\,\mathrm{K}$ but with a thinner sample because the original sample is too thick for Terahertz light to pass. The peak positions don't shift as we lower the temperature to $T = 1.6\,\mathrm{K}$, hence the g-factor doesn't change either (Fig. S5).

### 7. Fitting TDTS data

The spin-wave peak at each field and temperature in $\mathrm{Im}\tilde{\chi}(\nu)$ was fitted to a Lorentzian:

$$\mathrm{Im}\tilde{\chi}(\nu) \propto \frac{\frac{1}{2}\Delta}{(\nu - \nu_c)^2 + (\frac{1}{2}\Delta)^2},$$

to extract the resonant frequency $\nu_c$ and the FWHM width $\Delta$. A representative fit is show in Fig. S6**a**. The value of $\Delta$ as a function of temperature and field is shown in Fig. S6**b** and **c**, respectively. We find that $\Delta$ is mostly independent of temperature and field below $\sim 40\,\mathrm{K}$.

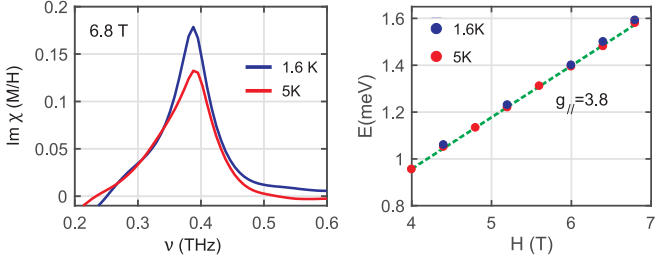
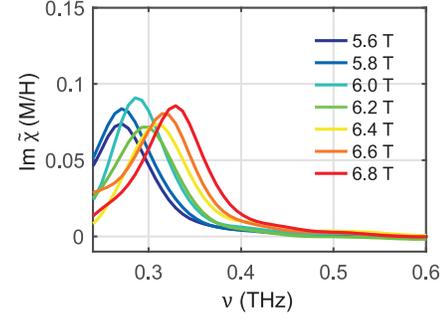

FIG. S5. **(a)** The imaginary part of the susceptibility $\mathrm{Im}\tilde{\chi}(\nu)$ at $+6.8\,\mathrm{T}$ for RCP light in the Faraday geometry taken at $1.6\,\mathrm{K}$ and $5\,\mathrm{K}$. **(b)** Spin wave excitation peaks versus magnetic fields at $1.6\,\mathrm{K}$ and $5\,\mathrm{K}$, showing the same g-factor.

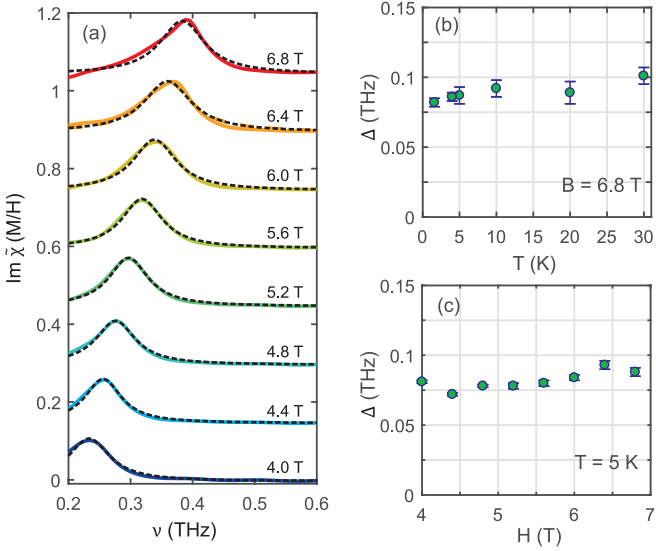

FIG. S6. **(a)** Representative Lorentzian fits to the TDTS data at $5\,\mathrm{K}$. The solid lines represent the imaginary susceptibility $\mathrm{Im}\tilde{\chi}(\nu)$ for RCP light from THz measurements in the Faraday geometry at different fields at $5\,\mathrm{K}$. The dashed lines represent the Lorentzian fit from which we extract the peak resonant frequency and linewidth. **(b)** Lorentzian linewidth $\Delta$ as a function of temperature $T$ from $1.6\,\mathrm{K}$ to $30\,\mathrm{K}$ at $H = 6.8\,\mathrm{T}$ and **(c)** as a function of field $H$ at $T = 5\,\mathrm{K}$. Error bars in **(b)** and **(c)** represent the 68% confidence interval (1 s.d.) in obtaining the fitting parameter $\Delta$.

## 8. TDTS data in the Voigt geometry

The TDTS spectra taken in the Voigt geometry is shown in Fig. S7. TDTS transmission specta were measured at $T = 5\,\mathrm{K}$ for various fields where the THz pulse $H_{dc} \perp c$ geometry. The spectra at each field are referenced to the spectra taken at $100\,\mathrm{K}$ for the same geometry. The resulting imaginary part of the magnetic susceptibility $\mathrm{Im}\tilde{\chi}(\nu)$ as calculated in the linear basis is shown in Fig. S7. The spin-wave peak resonant frequencies with field are plotted in Fig. 3e of the main text.

FIG. S7. The imaginary part of the susceptibility $\mathrm{Im}\tilde{\chi}(\nu)$ for linearly polarized light at different fields $H_{dc} \parallel a$ at $5\,\mathrm{K}$ as obtained by referencing the TDTS data to the spectra at $100\,\mathrm{K}$.

## 9. Spin-wave dispersions at $\mathbf{q} \neq 0$

Spin-wave dispersions in the field-saturated state for magnetic fields along the $c$ and $a$ axes are derived from the Hamiltonian in the main text, where $\mathbf{q}$ represents the spin-wave wave-vector in the first Brilloiun zone, and $\mathbf{a}_{1,i}$ and $\mathbf{a}_{2,i}$ represent the nearest and next-nearest neighbor vectors on the triangular-lattice, respectively, defined in Ref. [4],

$$
\begin{aligned}
E_c(\mathbf{q}) = \{ & [\mu_0 \mu_B g_\parallel H^z - 3(J_1^{zz} + J_2^{zz}) \\
& + 2\sum_{i=1}^{3} J_1^{\pm} \cos(\mathbf{q}\cdot\mathbf{a}_{1,i}) + 2\sum_{i=1}^{3} J_2^{\pm} \cos(\mathbf{q}\cdot\mathbf{a}_{2,i})]^2 \\
& - | 2J_1^{\pm\pm} \sum_{i=1}^{3} \gamma_{1,i}^{*} \cos(\mathbf{q}\cdot\mathbf{a}_{1,i}) |^2 \}^{1/2} 
\end{aligned} \quad (1)
$$

$$
\begin{aligned}
E_a(\mathbf{q}) = \{ & [\mu_0 \mu_B g_\perp H^x - 6(J_1^{\pm} + J_2^{\pm}) \\
& + \cos(\mathbf{q}\cdot\mathbf{a}_{1,1})(J_1^{\pm} + J_1^{zz}/2 - J_1^{\pm\pm}) \\
& + \cos(\mathbf{q}\cdot\mathbf{a}_{1,2})(J_1^{\pm} + J_1^{zz}/2 + J_1^{\pm\pm}/2) \\
& + \cos(\mathbf{q}\cdot\mathbf{a}_{1,3})(J_1^{\pm} + J_1^{zz}/2 + J_1^{\pm\pm}/2) \\
& + \sum_{i=1}^{3}\cos(\mathbf{q}\cdot\mathbf{a}_{2,i})(J_2^{\pm} + J_2^{zz}/2)]^2 \\
& - | \cos(\mathbf{q}\cdot\mathbf{a}_{1,1})(J_1^{\pm} - J_1^{zz}/2 - J_1^{\pm\pm} + iJ_1^{z\pm}) \\
& + \cos(\mathbf{q}\cdot\mathbf{a}_{1,2})(J_1^{\pm} - J_1^{zz}/2 + J_1^{\pm\pm}/2 - iJ_1^{z\pm}/2) \\
& + \cos(\mathbf{q}\cdot\mathbf{a}_{1,3})(J_1^{\pm} - J_1^{zz}/2 + J_1^{\pm\pm}/2 - iJ_1^{z\pm}/2) \\
& + \sum_{i=1}^{3}\cos(\mathbf{q}\cdot\mathbf{a}_{2,i})(J_2^{\pm} - J_2^{zz}/2) |^2 \}^{1/2} 
\end{aligned} \quad (2)
$$

## 10. $\chi^2$ goodness of fit and correlation plots

In this section we plot the $\chi^2$ goodness of fit maps for the global fit to the TDTS and INS data by varying $J_1^{\pm\pm}$ and $J_2/J_1$ (Fig. S8a) and by varying $J_1^{\pm\pm}$ and $|J_1^{z\pm}|$ (Fig. S8b). We also plot the correlation between each fitting parameters in global fit (Fig. S9).

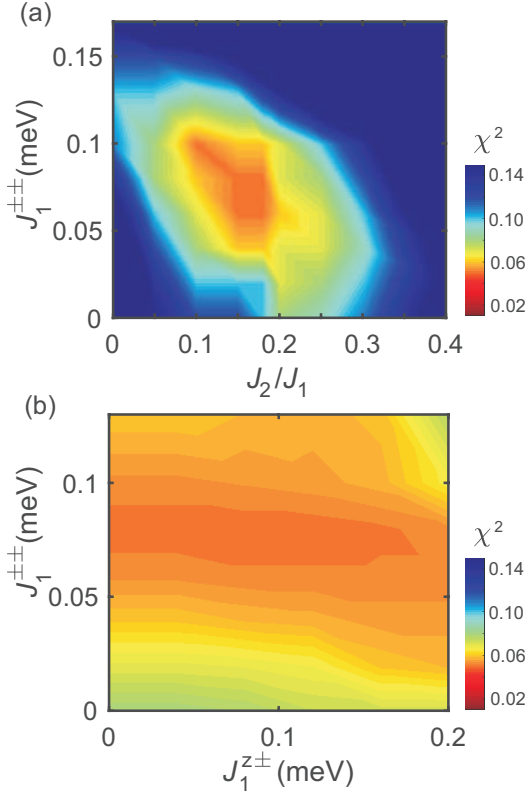

FIG. S8. Goodness of fit maps around the results of model C (Table I of main text) to indicate $\chi^2$ goodness of fit as a function of (a) $J_2/J_1$ vs $J_1^{\pm\pm}$ and (b) $|J_1^{z\pm}|$ vs $J_1^{\pm\pm}$ with unplotted parameters fixed to the values of model C.

## 11. TDTS linewidth calculations

Following the approach of Ref. [5] and noting that TDTS functions as high-field ESR, we calculate the dependence of the linewidth of the TDTS spin-wave peaks on the anisotropic exchange using the expression $\Delta_{cal} = \sqrt{2\pi M_2^3/M_4}$, where $M_2 = \langle [\mathcal{H}', M^+][M^-, \mathcal{H}']\rangle/\langle M^+M^-\rangle$ is the second moment and $M_4 = \langle [\mathcal{H},[\mathcal{H}', M^+]][\mathcal{H},[M^-, \mathcal{H}']]\rangle/\langle M^+M^-\rangle$ is the fourth moment. Here, the the anisotropic part $\mathcal{H}'$ can be obtained by subtracting the isotropic part $J_0 = (4J_\pm + J_{zz})/3$ from the spin Hamiltonian with $J_2 = 0$. This is a similar analysis to that of the previous ESR work on YbMgGaO$_4$ [6], but performed here in the high field regime. The full expressions for $M_2$ and $M_4$ with $J_2 = 0$ are given explicitly in the SI of [6]. Note that it was pointed out previously that both hyperfine and dipolar interactions are too small to cause the broadening.

Recall that model B* gives only a slightly worse fit to the spin-wave data than model C, as can been seen in Fig. 4 of the main text. To distinguish between the two, we determine the discrepancy between the calculated and the experimental linewidths in both the Faraday $\Delta_f$ and Voigt $\Delta_v$ geometry as follows: we define a function $R_p = \frac{1}{2}[|\frac{\Delta_f - \Delta_{cal}}{\Delta_f}| + |\frac{\Delta_v - \Delta_{cal}}{\Delta_v}|]$. Fig. 5 in the main text shows $R_p$, calculated using the expressions for $\Delta$, $M_2$ and $M_4$, for the case of model B* ($J_2 = 0$) as a function of $|J_1^{z\pm}|$ and $|J_1^{\pm\pm}|$. $R_p$ is minimized along the yellow contour, and hence within a model where $J_2 = 0$ one expects the anisotropic exchange parameters to fall in this yellow region. However, note that the model with $J_2 = 0$ only fits the neutron and THz data in the global fit if $J_1^{\pm\pm}$ is about 0.13(2) meV (model B* in Table I of the main text). But if $J_1^{\pm\pm}$ was as large as 0.13(2) meV with $J_2 = 0$, then a much broader linewidth in the THz data would have been observed. This gives further evidence that model B*, with no next-nearest neighbor interaction is inappropriate.

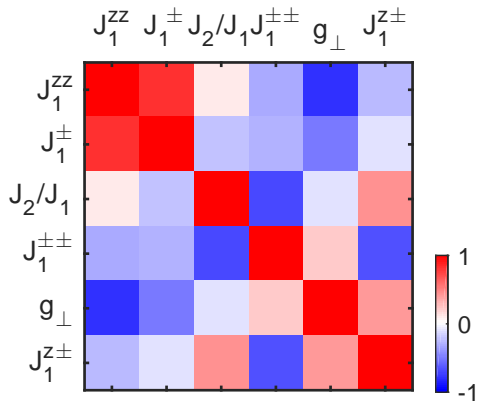

FIG. S9. The correlation plot of six target parameters